\documentclass[amsmath,twocolumn,aps,prb]{revtex4}
\usepackage{bm}
\usepackage{color}
\usepackage{graphicx}
\usepackage{bbold}
\newcommand{\bq}{\begin{equation}}
\newcommand{\ee}{\end{equation}}
\begin{document}

\title{Dynamic response functions of two-dimensional Dirac fermions with screened Coulomb and short-range interactions}

\author{M. Agarwal}
\affiliation{Department of Physics and Astronomy, University of Utah, Salt Lake City, UT 84112, USA}
\author{E. G. Mishchenko}
\affiliation{Department of Physics and Astronomy, University of Utah, Salt Lake City, UT 84112, USA}

\begin{abstract}
We consider a screened Coulomb interaction between electrons in graphene and determine their dynamic response functions, such as a longitudinal and a transverse electric conductivity and a polarization function and compare them to the corresponding quantities in the short-range interaction model.
The calculations are performed to all orders for short-range interaction by taking into account the self-energy renormalization of the electron velocity and using a ladder approximation to account for the vertex corrections, ensuring that the Ward identity (charge conservation law) is satisfied. Our findings predict a resonant response of interacting electron-hole pairs at a particular frequency below the threshold $qv=\omega$ and further predict an instability for sufficiently strong interactions.
\end{abstract}

\maketitle
\section{Introduction}

Response functions of a physical system are measurable quantities that contain a variety of valuable information about spectra and interactions of the system's quasiparticles.\cite{PN} For example, the polarization function $\Pi(\omega,{\bf q})$ determines a density response to a scalar field oscillating with frequency $\omega$ and wave vector ${\bf q}$. Other response functions include dielectric function $\epsilon (\omega, {\bf q})$, which determines screening of external electric field; dynamic conductivity $\sigma_{jk}(\omega, {\bf q})$, which determines a response to an electric field; spin susceptibility, which determines a response to a magnetic field; and so on. Some of the functions obey relationships that result from various symmetries, such as the time-reversal symmetry, charge conservation, and the like. For example, as a result of the charge conservation, the longitudinal conductivity $\sigma_{xx}(\omega, {\bf q})$ is proportional to the polarization function $\Pi(\omega,{\bf q})$.
Response functions of two-dimensional Dirac fermions, such as electrons in graphene,\cite{CN} have been studied extensively. The polarization function of the intrinsic (disorder-free undoped) graphene at zero temperature is\cite{Shung,Gonzales}
\begin{equation}
\label{Non-interacting polarization}
\Pi(\omega,q) = -\frac{Nq^2}{16\sqrt{q^2v_0^2 - \omega^2}},
\end{equation}
where $N$ is the number of fermion species and $v_0$ is the band Dirac velocity (we set $\hbar =1$ throughout this paper). In graphene, $N=4$, due to the spin and valley degeneracy. The expression (\ref{Non-interacting polarization}) indicates (as explained below) that the uniform ($q=0$) conductivity\cite{LFS} is $\sigma(\omega,0) =Ne^2/16\equiv \sigma_0$, independent of the frequency $\omega$. This value has indeed been observed in optical experiments.\cite{NBG}

Electron-electron interactions were understood\cite{SS1,HJV} to cause a departure of the conductivity from the value $\sigma_0$. Surprisingly, to the first order in the interaction strength $g=e^2/v_0$, the numerical value of the interaction correction is rather small,\cite{M,SS2,AVP,SF,TK,LOS} $\sigma(\omega)=\sigma_0(1+0.01 e^2/v_0)$. This is remarkable since the electron-electron interaction in intrinsic graphene is not weak. Indeed, the dielectric function, which in the mean-field (random phase) approximation is given by $\epsilon (\omega, {\bf q})=1-2\pi e^2 \Pi(\omega,{\bf q})/q$, amounts in the static limit $\omega=0$ to a mere constant, $\epsilon = 1+\pi N e^2/8v\approx 4.4$, which does not change the long-range form of the Coulomb interaction.
The almost complete disappearance of the interaction corrections to the conductivity occurs as a result of a peculiar cancelation of the self-energy and the vertex corrections.

Calculations of finite-${q}$ interaction corrections to the conductivity, as well as the  higher order ($\sim e^4$) corrections, are rather challenging. Even in the homogeneous ($q=0$) limit, it remains unknown if the weak sensitivity to the interaction strength persists beyond the first order in $e^2$.
In the present paper, we consider how the response functions of graphene for non-zero ${q}$ are affected by a screened Coulomb interaction. We assume that the screening is of a type produced by a conducting gate located at distance $d/2$ from the plane of graphene. The Coulomb interaction in the momentum space is then
 \begin{equation}
 \label{screened interaction}
U_q = \frac{U_{0}}{qd} \left(1-e^{-qd}\right),~~~ U_{0} = 2\pi e^2 d.
\end{equation}
The second term in this expression describes interaction with an image charge induced on the gate.
At large momenta, $qd \gg 1$ (short distances), the interaction assumes the usual unscreened Coulomb form $U_q \to  {2\pi e^2}/{q}$, and for small momenta, $qd \ll 1$ (long distances), it tends to a constant: $U_q \approx U_{0}$. Some aspects of a fully short-range interaction, where $U_q\equiv U_0$ for all $q$, have been previously considered.\cite{JVH,GMP,AVP} The screened interaction (\ref{screened interaction}) allows, on one hand, to collect logarithmic contributions from large momenta, while, on the other, simplifies calculations of the convergent small-momenta integrals (compared with the more difficult situation of the unscreened Coulomb interaction).

This paper is organized as follows: In Sec.~II, the self-energy (electron velocity renormalization) for the interaction (\ref{screened interaction}) is calculated. In Sec.~III, the first order interaction correction to the polarization function is determined for arbitrary $\omega$ and $q$. In Sec. IV, a similar calculation is carried out for the first order current-current correlation function. It is shown that the longitudinal conductivity and the polarization function satisfy a relation that follows from the charge conservation, provided that a proper ultraviolet regularization procedure is implemented. In Sec.~V, we perform calculations beyond the first order perturbation theory by summing up an infinite series of ladder diagrams. Such approach becomes progressively more accurate in the vicinity of the threshold for electron-hole pair production, $\omega \approx qv$. Using the results of the ladder summation, we show that the response of graphene becomes resonant at a specific frequency.

\section{The Electron Self-Energy}

We consider the following Hamiltonian describing two-dimensional Dirac fermions,
\begin{equation}
\label{hamil}
H = v_{0}\sum_{\bm{p}} \hat c_{\bf{p}}^{\dagger} \hat{\bm \sigma} \cdot {\bf p} \hat c_{{\bf p}} + \frac{1}{2} \sum_{{\bf p,k,q}} U_q (\hat c_{\bf{p-q}}^{\dagger}\hat c_{{\bf p}}) (\hat c_{{\bf k+q}}^{\dagger}  \hat c_{{\bf k}}) ,
\end{equation}
where ``hats'' denote operators in the pseudo-spin (sublattice) space and $\hat {\bm \sigma}$ stand for the set of  Pauli matrices in that space. The parentheses indicate inner products of pseudo-spinors. The usual spin summation, as well as the summation over fermion flavors (multiple Dirac points) is assumed to be performed in the Hamiltonian (\ref{hamil}). The summation over momenta in Eq.~(\ref{hamil}) and throughout the paper is understood as the two-dimensional integral, $\sum_{\bm{p}} \equiv \int d^2p/(2\pi)^2$, with the normalization volume (area) set to $1$.
The Hamiltonian (\ref{hamil}) is known to describe the low-energy properties of a monolayer graphene. We also assume that the interaction $U_q$ is sufficiently weak at large $q$ and does not cause transitions between different Dirac points.

The first order   correction in $U_q$ to the polarization function (\ref{Non-interacting polarization}) is shown in Fig.~1b). Solid lines correspond to time-ordered electron Green's functions,  which in the energy-momentum representation (and at zero temperature),
\begin{equation}
\label{Green's function}
\hat G_{\epsilon, {\bf k}} = \frac{1}{2}\sum_{\beta}\frac{1+\beta \hat\sigma_{{\bf k}}}{\epsilon-\beta(v_{0}k - i\eta)},
\end{equation}
are given by a sum over the conduction subband $\beta=1$ (the upper Dirac cone) and the valence subband $\beta =-1$ (the lower Dirac cone). The operator $\hat\sigma_{{\bf k}} = \hat {\bm \sigma} \cdot {\bf n}$ is the projection of the pseudospin operator onto the direction of the electron momentum ${\bf n} = {\bf k}/k$.
\begin{figure}
\includegraphics[width=6cm,height=6cm,keepaspectratio]{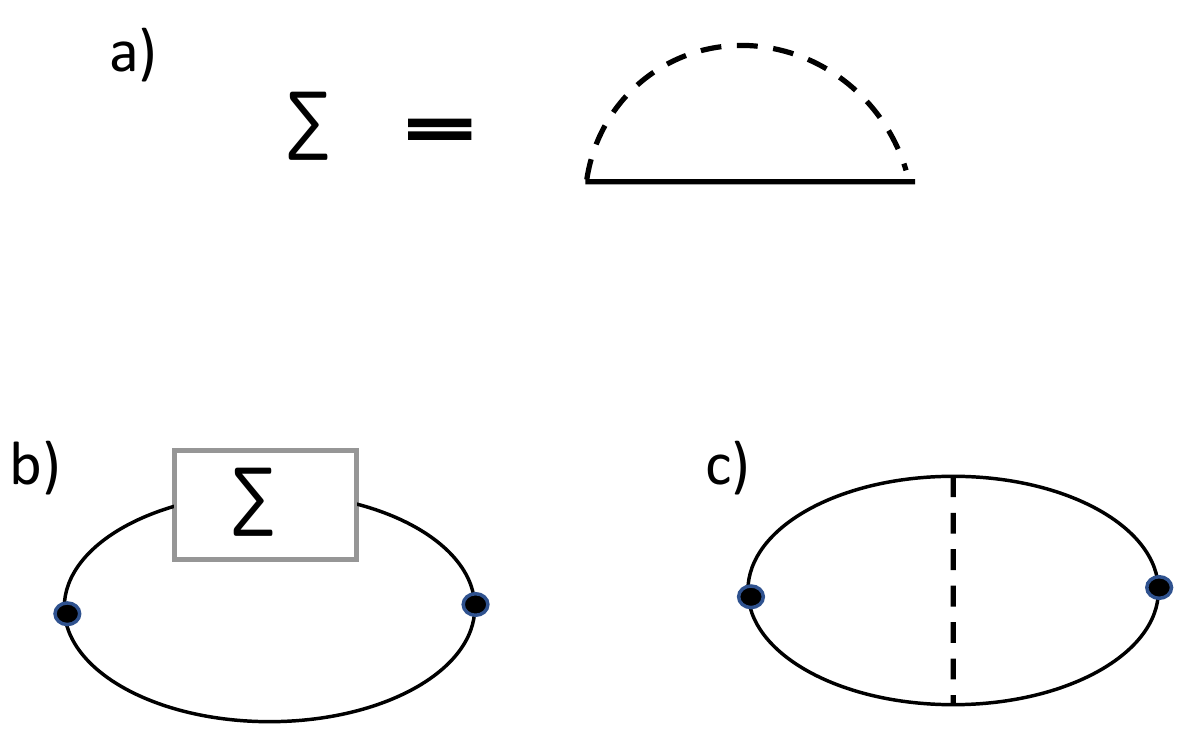}
\caption{a) First order self-energy diagram b) self-energy, and c) vertex correction to the polarization function and conductivity to first order in interaction. The vertex equals 1 for polarization function and $ev_{0} \bm{\hat\sigma}$ for conductivity.}
\label{Fig. 1}
\end{figure}

The electron self-energy, illustrated in Fig.~1a),  is
\begin{equation}
\label{Self-energy1}
{\textstyle \hat \Sigma_ {\bf p}} = i\sum_{{\bf k}}U_{|{\bf p}-{\bf k}|} \int \frac{d\epsilon}{2\pi}  \hat G_{\epsilon, {\bf k}} e^{i\epsilon \, 0+}.
\end{equation}
Formally, the momentum integral here has a power-law divergence. This divergence simply amounts to a renormalization of the Fermi energy; the much smaller difference $\Sigma_ {\bf p} \to \Sigma_ {\bf p}-\Sigma_{0}$ results in the electron velocity renormalization,
\begin{align}
\label{velocity renormalization 0}
\Sigma_{\bf p}-\Sigma_{0}  = \frac{1}{2} \sum_{{\bf k}}U_{|{\bf p}-{\bf k}|}\hat\sigma_{{\bf k}}.
\end{align}
From the isotropy of the system, it follows that the self-energy is $\Sigma_ {\bf p}-\Sigma_{0} = p \delta v_p \hat \sigma_{\bf p}$, with the velocity correction being,
\begin{equation}
\label{velocity renormalization 1}
\delta v_p = \frac{1}{2p} \sum_{\bf k}U_{|{\bf p}-{\bf k}|} \cos{\theta},
\end{equation}
where $\theta$ is the angle between ${\bf k}$ and ${\bf p}$. The integral in Eq.~(\ref{velocity renormalization 1}), if non-zero, arises from momenta $k \gg p$. It is, therefore, sufficient to determine the integral to the linear order in the external momentum $p$. Namely, using $|{\bf p}-{\bf k}| \approx k-p\cos\theta$, and noticing that the $p$-independent term vanishes because of the angle integral, we arrive at,
\begin{equation}
\label{velocity renormalization 2}
\delta v =- \frac{1}{8\pi} \int\limits_0^\infty  kdk \frac{\partial U}{\partial k}.
\end{equation}
This formula implies the absence of a low-$k$ singularity in the integrand; otherwise (as in the case of the unscreened Coulomb interaction), the lower limit cannot be extended to $k=0$.

The {\it screened Coulomb}  interaction (\ref{screened interaction}) leads to a logarithmically divergent (at large $k$) integral in Eq.~(\ref{velocity renormalization 2}). It can be regularized, for example, by replacing $U_q \to U_q e^{-q/\Lambda}$,
\begin{align}
\label{velocity renormalization 3}
\delta v =& \frac{e^2}{4} \int\limits_0^\infty \frac{dk}{k} \left( 1- e^{-kd} \right) e^{-k/\Lambda}\nonumber\\
			   & =\frac{e^2}{4} \ln{( \Lambda d)} +O\left(\frac{1}{\Lambda d} \right).
\end{align}
Alternatively, the integral in Eq.~(\ref{velocity renormalization 2}) can be extended to the upper limit $k=\Lambda$. This only changes the numerical coefficient under the logarithm in Eq.~(\ref{velocity renormalization 3}). Since $\Lambda$ is only known by an order of magnitude (being of the order of the inverse interatomic distance), the two approaches lead to essentially the same result for the velocity renormalization.

The derived result for $\delta v$ is valid provided that $p\ll 1/d$. Compared with the unscreened Coulomb interaction case, the curvature of the electron spectrum is absent. For higher values of the electron momenta,  $1/d \ll p \ll \Lambda$, Eq.~(\ref{velocity renormalization 2}) reproduces the usual velocity correction,\cite{Gonzales} $\delta v =\frac{e^2}{4} \ln{(\Lambda/p)}$, after the lower limit is replaced with $p$.

In the case of a {\it short-range} interaction, where $U=U_0$ for all momenta, the independence of the general expression (\ref{velocity renormalization 2}) of the momentum ${p}$ predicts that $\delta v=0$.

\section{The first order correction to the polarization function}

Turning now to the interaction corrections, we first consider the polarization function, which is the (time-ordered) density-density correlation function,
\begin{equation}
\Pi(t,{\bf q}) =-i\sum_{\bf k} \langle T \rho(t,{\bf q}) \rho(0,{\bf k})\rangle,
\end{equation}
where
\begin{equation}
 \rho(t,{\bf q}) =\sum_{\bf p} (c_{\bf p}^\dagger (t) c_{{\bf p}+{\bf q}} (t))
\end{equation}
is the electron density operator (with the summation over the fermion species implied). In the frequency representation, the vertex correction,  given by the diagram in Fig.~\ref{Fig. 1}c), amounts to the  integral,
\begin{align}
\label{first order vertex}
\Pi^{(1)}_{V}(\omega,q) &=  N \text{Tr} \sum_{\bf{p},\bf{p'}} U_{|{\bf p-p'}|}\int \frac{d\epsilon }{2\pi}\hat G_{\epsilon_+ \, {\bf p}_+} \hat G_{\epsilon_- \,{\bf p}_-} \nonumber\\
						&\times \int \frac{d\epsilon' }{2\pi} \hat G_{\epsilon'_-\, {\bf p'}_- } \hat G_{\epsilon'_+ \,{\bf p'}_+},
\end{align}
where $\epsilon_\pm =\epsilon\pm \omega/2$ and ${\bf p}_\pm ={\bf p}\pm{\bf q}/2$.
The energy integrals (calculated in the Appendix) are
\begin{align}
\label{useful integral}
&\int \frac{d\epsilon }{2\pi}\hat G_{\epsilon_+\, {\bf p}_+} \hat G_{\epsilon_- \,{\bf p}_-}\nonumber \\ &= i\frac{ \Omega_{\bf p}(1-\hat\sigma_{{\bf p}_+} \hat\sigma_{{\bf p}_-}) +\omega(\hat\sigma_{{\bf p}_+} - \hat\sigma_{{\bf p}_-}) }{2(\omega^2-\Omega^2_{\bf p})},
\end{align}
where $\Omega_{\bf p} = v_0 (p_++p_-)$ is the energy of an electron-hole pair.
(The integral involving the primed quantities in Eq.~(\ref{first order vertex}) has the identical value.)
The singularity in the right-hand side of Eq.~(\ref{first order vertex}) (as well as in similar integrals encountered below) should be understood to be regularized by shifting it away from the real frequencies: $\omega^2 \to \omega^2 +i0$ for the time-ordered polarization function, and $\omega^2 \to (\omega +i0)^2$ for the  retarded function. (The two functions coincide when $\omega>0$.)

It follows from the form of the last expression (which decreases sufficiently fast for large $p,\,p' \gg q$) that the remaining momentum integral over ${\bf p}$ (and, similarly, the integral over ${\bf p}'$) converges over characteristic momenta $\sim q, \omega/v_0$ that are assumed to be much smaller than $1/d$.
This indicates that the interaction can be approximated with its zero-momentum value, $U_{|{\bf p-p'}|} \approx U_0$.
Accordingly, the remaining momentum integrals over ${\bf p}$ and ${\bf p}'$  decouple and can be calculated exactly\cite{footnote1} (see Appendix):
\begin{align}
\label{useful integral 1}
&\sum_{\bf{p}} \int \frac{d\epsilon }{2\pi}\hat G_{\epsilon_+\, {\bf p}_+} \hat G_{\epsilon_- \,{\bf p}_-}=-\frac{iq(qv_{0}+\omega \hat\sigma_{\bf q})}{32v_{0}\sqrt{q^2v_{0}^2-\omega^2}}.
\end{align}
As a result, the first-order vertex correction (\ref{first order vertex}) becomes,
\begin{equation}
\label{Polarization vertex correction}
\Pi^{(1)}_{V}(\omega,q) = -U_{0} \frac{Nq^2(\omega^2+q^2v_{0}^2)}{2(16v_{0})^2(q^2v_{0}^2-\omega^2)}.
\end{equation}
We note that this result has been previously obtained in Ref.~\onlinecite{JVH}.

The first-order self-energy correction can be obtained directly from the zero-order polarization function (\ref{Non-interacting polarization}) by noticing that in the case of the screened Coulomb interaction and $q \ll 1/d$, the velocity renormalization (\ref{velocity renormalization 3}) does not introduce any spectrum curvature for relevant momenta. This means that it is sufficient to replace $v_0 \to v_0 +\delta v$ in the zero-order expression (\ref{Non-interacting polarization}) and then expand it to the lowest order in $\delta v$. This gives
\begin{equation}
\label{Polarization self-energy correction}
 \Pi^{(1)}_{SE}(\omega, q) = \frac{N q^4 v_{0}\delta v}{16 (q^2v_{0}^2-\omega^2)^{3/2}}.
\end{equation}

In the case of the short-range interaction, $\delta v=0$ and the self-energy correction (\ref{Polarization self-energy correction}) is absent. The absence of velocity renormalization seems to be consistent with recent numerical calculations.\cite{TLR}

Comparing the value of the vertex correction to the self-energy correction, we observe that the vertex correction dominates for small wave vectors $q$. In contrast, the self-energy correction becomes more important at  $\omega \approx qv_{0}$. This is not surprising since the velocity renormalization results in a shift of the singular edge of the electron-hole continuum.
Indeed, the line $\omega = qv_0$ separates the region ($\omega > qv_0$) where electron-hole pairs can be generated by a field with frequency $\omega$ and wave vector $q$, from the region ($\omega < qv_0$) where no real electron-hole pairs are excited. In the former region, the zero-order polarization function (\ref{Non-interacting polarization}) is purely imaginary, while in the latter region it is purely real. The choice of the branch of the square root is always such that the imaginary part of the polarization function is negative for positive $\omega$.

\section{The first order correction to the current-current correlation function}

To test whether an interacting theory makes sensible predictions, one has to verify that the theory respects charge conservation. The latter ties a response of the system to an electrostatic scalar potential to a response to the related vector electric field.

Suppose that a time- and position-dependent  electrostatic potential $\phi(\omega,{\bf q})$ acts in the system. It causes a charge density variation that is determined by the polarization function: $\delta \rho (\omega,{\bf q}) =e^2 \Pi (\omega,{\bf q})\phi(\omega,{\bf q})$. According to the charge conservation, this charge density must be accompanied by an electric current ${\bf j}(\omega,{\bf q})$ that satisfies the continuity equation, $\omega \rho (\omega,{\bf q})  ={\bf q}\cdot {\bf j}(\omega,{\bf q})$. This determines the {\it longitudinal} part (parallel to ${\bf q}$) of the current: ${\bf j}_\parallel (\omega,{\bf q}) ={\bf q} \omega \rho (\omega,{\bf q})/q^2$. But the same current can be considered as a response to the electric field ${\bf E}(\omega,{\bf q}) =-i{\bf q} \phi (\omega,{\bf q})$, determined by the system's conductivity:   ${\bf j}(\omega,{\bf q}) = \sigma_\parallel (\omega,{\bf q}) {\bf E}(\omega,{\bf q})$. By comparing the two results, we can conclude that the two response functions must be related as,
\begin{equation}
\label{Ward identity}
\sigma_\parallel (\omega,{\bf q}) =\frac{ie^2 \omega}{q^2} \Pi (\omega,{\bf q}).
\end{equation}
No such relation exists for the transverse conductivity $\sigma_\perp(\omega,{\bf q})$, which determines a response of the system to electric field that is perpendicular to ${\bf q}$. For an isotropic system (considered here) the conductivity tensor is given by,
\begin{equation}
\sigma_{jk} (\omega,{\bf q}) = \frac{q_j q_k}{q^2} \sigma_\parallel (\omega,q) +\left(\delta_{jk} - \frac{q_j q_k}{q^2} \right) \sigma_\perp (\omega,q).
\end{equation}

In the Kubo approach, the electric conductivity is calculated from the current-current correlation function. Because the  operator of electric current of the Dirac fermions is proportional to their pseudospin, ${\bf j} = e v_0 {\bm \sigma}$,
the conductivity is determined by the correlator,
\begin{equation}
\Pi_{jk}(t,{\bf q}) =-i\sum_{\bf k} \langle T \rho_j(t,{\bf q}) \rho_k(0,{\bf k})\rangle,
\end{equation}
of the pseudospin density,
\begin{equation}
 \rho_j(t,{\bf q}) =\sum_{\bf p} (c_{\bf p}^\dagger (t) \hat \sigma_j c_{{\bf p}+{\bf q}} (t)).
\end{equation}
The conductivity, in turn, is determined by the pseudospin density correlator:
\begin{equation}
\label{Kubo formula}
\sigma_{jk} (\omega,{\bf q}) =i \frac{e^2v_0^2}{\omega}\Pi_{jk}^{(0)}   (\omega,{\bf q}).
\end{equation}
 Note that in terms of $ \Pi_{jk}^{(0)}  (\omega,{\bf q})$, the charge conservation condition (\ref{Ward identity}) can also be represented in the equivalent form,
\begin{equation}
\label{Ward identity1}
\Pi_{xx}^{(0)}  (\omega,{\bf q}) =\frac{\omega^2}{q^2v_0^2} \Pi (\omega,{\bf q}),
\end{equation}
where $x$ is the direction of the wave vector, ${\bf q}=q\hat{\bf x}$.

\subsection{The Self-Energy Correction}

To the zeroth order in the interaction, the longitudinal pseudospin density correlator is
\begin{align}
\label{Kubo0}
\Pi_{xx}^{(0)}  (\omega,{\bf q}) = -i N \text{Tr} \sum_{\bf p}\int \frac{d\epsilon }{2\pi}\hat G_{\epsilon_+\, {\bf p}_+} \hat\sigma_{x} \hat G_{\epsilon_- \,{\bf p}_-} \hat \sigma_x.
\end{align}
The energy integral here is taken similarly to Eq.~(\ref{useful integral 1}),
\begin{align}
\label{useful integral2}
&\int \frac{d\epsilon }{2\pi}\hat G_{\epsilon_+\, {\bf p}_+} \hat\sigma_{x} \hat G_{\epsilon_- \,{\bf p}_-}\nonumber \\
&= i\frac{\omega  (\hat\sigma_{{\bf p}_+}\hat\sigma_{x} - \hat\sigma_{x}\hat\sigma_{{\bf p}_-}) +\Omega_{\bf p}(\hat\sigma_{x}-\hat\sigma_{{\bf p}_+}\hat\sigma_{x} \hat\sigma_{{\bf p}_-})}{2(\omega^2-\Omega_{\bf p}^2)}.
\end{align}
Multiplying this expression by $\hat \sigma_x$ and taking the trace, we obtain,
\begin{equation}
\label{kubo zero order}
\Pi_{xx}^{(0)}  (\omega,{\bf q})= N \sum_{\bf p} \frac{\Omega_{\bf p} A_{\bf p}}{\omega^2-\Omega_{\bf p}^2},
\end{equation}
where $A_{\bf p}= 1-\cos(\theta_{{\bf p}_+}+\theta_{{\bf p}_-})$, and $\theta_{{\bf p}_\pm}$ denote the angles that the momenta ${\bf p}_\pm$ make with the direction of ${\bf q}$.

The obtained integral is power-law divergent and predicts that the conductivity (\ref{Kubo formula}) is dependent on the cut-off $\Lambda$, in violation of the charge conservation (\ref{Ward identity}) whose right-hand side, determined by Eq.~(\ref{Non-interacting polarization}), is cut-off independent. One way to regularize this divergence is to subtract from the integrand in Eq.~(\ref{kubo zero order}) its zero-frequency value. This results in the following expression (see Appendix for details),
\begin{align}
\label{zero-order conductivity}
\Pi_{xx}^{(0)}  (\omega,{\bf q}) \to & N{\omega^2} \sum_{\bf p} \frac{ A_{\bf p}}{\Omega_{\bf p}(\omega^2-\Omega_{\bf p}^2)}\nonumber\\ &= -\frac{N\omega^2}{16v_0^2 \sqrt{q^2v_0^2 - \omega^2}},
\end{align}
which is consistent with the charge conservation.

The same outcome occurs if one uses dimensional regularization.\cite{JVH} Let us illustrate the equivalency of the two approaches using the limit of $q=0$ as an example, in which case the angular integral ensures that $A_{\bf p}$ can be replaced with unity. The remaining momentum integral is ($a=\omega/v_0$)
\begin{align}
\label{Int_beta}
I_\beta =\int\limits_0^\infty \frac{p^\beta\,dp }{a^2-p^2+i0},
\end{align}
with $\beta=2$. The dimensional regularization approach entertains non-integer number of space-time dimensions and ultimately amounts to calculating the last integral in the range $-1<\beta < 1$, where it is convergent, and then analytically continuing it outside this range, e.g., to $1<\beta < 3$.

Since the pole of the integrand is right below the real axis, we can rotate the integration path until it follows the positive half of the imaginary axis, $p=iy$:
\begin{align}
\label{dimensional regularization}
I_\beta =&\int\limits_0^\infty \frac{p^\beta\,dp }{a^2-p^2+i0}= i^{\beta +1}
\int\limits_0^\infty \frac{y^\beta\,dy }{a^2+y^2}\nonumber\\ &=i^{\beta +1} \frac{\pi a^{\beta-1}}{2 \cos{(\frac{\pi \beta}{2})}}.
\end{align}
Analytical continuation of the obtained result to the range $1<\beta < 3$ yields, $I_{\beta \to 2} = i\pi a/2$. It is now easy to verify that the subtraction of the $a=0$ value from the integrand in Eq.~(\ref{Int_beta}) leads to the same result. Indeed, working this time directly in the range $1<\beta < 3$ of interest, we write upon the subtraction:
\begin{align}
I_\beta =&\int\limits_0^\infty dz \left[ \frac{p^\beta}{a^2-p^2+i0} +p^{\beta-2} \right]\nonumber\\ &= a^2
 \int\limits_0^\infty  \frac{p^{\beta-2}\, dp}{a^2-p^2+i0} = i^{\beta-1}a^2
\int\limits_0^\infty \frac{y^{\beta-2}\,dy }{a^2+ y^2}\nonumber\\ &=i^{\beta-1}\frac{\pi a^{\beta-1}}{2 \cos{(\frac{\pi (\beta-2)}{2})}}=-i^{\beta-1}\frac{\pi a^{\beta-1}}{2 \cos{(\frac{\pi \beta}{2})}},
\end{align}
which coincides with Eq.~(\ref{dimensional regularization}).

With the help of the zero-order conductivity (\ref{zero-order conductivity}), the first-order self-energy correction can now be obtained in the same way as used above to obtain the self-energy contribution to the polarization function -- by replacing  $v_0 \to v_0 +\delta v$  and expanding to the first order in $\delta v$:
\begin{equation}
\label{self-energy Kubo}
 (\Pi_{xx})^{(1)}_{SE}  (\omega,{\bf q}) = \frac{N\omega^2}{16v^3_{0}}\frac{(3q^2v_{0}^2-2\omega^2)}{(q^2v_{0}^2-\omega^2)^{3/2}} \delta v.
\end{equation}
This result should be added to the first-order vertex correction, which is calculated in the following subsection.

\subsection{The Vertex Correction}

Let us turn to the first-order, in the screened Coulomb interaction (\ref{screened interaction}), vertex correction to the current-current correlation function given by the diagram in Fig.~\ref{Fig. 1}c) (where now the vertices are the pseudospin matrices $\hat \sigma_x$),
\begin{align}
\label{first order vertex conductivity}
(\Pi_{xx}^{(1)})_{V}  (\omega,{\bf q})  = &N \text{Tr}\sum_{\bf{p},\bf{p'}} U_{|{\bf p-p'}|}\int \frac{d\epsilon }{2\pi}\hat G_{\epsilon_+ \, {\bf p}_+} \hat\sigma_{x} \hat G_{\epsilon_- \,{\bf p}_-} \nonumber\\
						&\times \int \frac{d\epsilon' }{2\pi} \hat G_{\epsilon'_-\, {\bf p'}_- } \hat\sigma_{x} \hat G_{\epsilon'_+ \,{\bf p'}_+}.
\end{align}
The energy integrals here are the same as in Eq.~(\ref{useful integral2}) (the integrals involving primed and unprimed quantities have the same value).
Performing the trace operation, we write,
\begin{equation}
\label{Ver_cond1}
(\Pi_{xx}^{(1)})_{V}  =\frac{-N}{2} \sum_{\bf{p},\bf{p'}}U_{|\bf{p}-\bf{p'}|}\frac{\Omega_{\bf p}\Omega_{{\bf p}'}A_{\bf p}A_{{\bf p}'}+\omega^2 B_{\bf p}B_{{\bf p}'}}{(\omega^2-\Omega_{\bf p}^2)(\omega^2-\Omega^2_{{\bf p}'})},
\end{equation}
where we denote $B_{\bf p}= \cos{\theta_{{\bf p_+}}} - \cos{\theta_{{\bf p_-}}}$.
The integrals in Eq.~(\ref{Ver_cond1}) are formally divergent. To regularize them, we subtract from the integrand its value at zero frequency,
\begin{align}
\label{Ver_cond2}
(\Pi_{xx}^{(1)})_{V}   =& \frac{-N\omega^2 }{2} \sum_{\bf{p},\bf{p'}}U_{|\bf{p}-\bf{p'}|} \nonumber\\
				   & \times \frac{ (\Omega_{\bf p}^2+\Omega_{{\bf p}'}^2- \omega^2)A_{\bf p}A_{{\bf p}'} +\Omega_{\bf p}\Omega_{{\bf p}'}B_{\bf p}B_{{\bf p}'}  }{\Omega_{\bf p}\Omega_{{\bf p}'}(\omega^2-\Omega_{\bf p}^2)(\omega^2-\Omega^2_{{\bf p}'})}.
\end{align}
Although the subtraction of the zero frequency value removes the power-law singularity, the integrals in Eq.~(\ref{Ver_cond2}) still lead to a logarithmic divergence at large  $p,p'$.
 Because $\Omega_{\bf p}$ is large at large momenta, $B_{\bf p}$ is small there, and $A_{\bf p} \to 1$, the logarithmic contribution originates with the term $(\Omega_{\bf p}^2+\Omega_{{\bf p}'}^2)A_{\bf p}A_{{\bf p}'}$. Since the integral is symmetric with respect to ${\bf p}$ and ${\bf p}'$, we can rewrite the singular part of the $A_{\bf p}A_{{\bf p}'}$ contribution as,
\begin{align}
\label{Ver_cond3}
(\Pi_{xx}^{(1)})_{V-AA} = - N\omega^2 \sum_{\bf{p}}  \frac{\Omega_{\bf p}A_{\bf p} }{\omega^2-\Omega_{\bf p}^2}
\sum_{\bf{p'}}
						 \frac{U_{|{\bf p}-{\bf p'}|}A_{{\bf p}'}}{\Omega_{{\bf p}'}(\omega^2-\Omega^2_{{\bf p}'}) }
\nonumber\\ =
N\omega^2 \sum_{\bf{p}}  \left(  \frac{A_{\bf p} }{\Omega_{\bf p}} -\frac{\omega^2 A_{\bf p}}{\Omega_{\bf p} (\omega^2-\Omega_{\bf p}^2)} \right)\sum_{\bf{p'}}
\frac{U_{|{\bf p}-{\bf p'}|}A_{{\bf p}'}}{\Omega_{{\bf p}'}(\omega^2-\Omega^2_{{\bf p}'}) }.
\end{align}

The first term in the parenthesis leads to a divergent $p$-integral. Note that because the $p'$ integral is convergent, the singular contribution arises from such momenta that $p \gg p'$,
where we can approximate  $U_{|{\bf p-p'}|} \approx U_p$. In the resulting $p$ integral, we notice that at large momenta $p$ the integrand, $A_{\bf p} /{\Omega_{\bf p}} \to 1/(2v_0p)$. We can thus write the $p$ integral, by separating this singular contribution, as
\begin{equation}
\label{vertex singular}
\sum_{\bf{p}} \left( \frac{A_{\bf p} }{\Omega_{\bf p}} -\frac{1}{2v_0p}\right)U_p +\frac{1}{4\pi v_{0}} \int\limits_0^\infty dp \,U_p.
\end{equation}
The first integral is convergent over $p\sim q$. Because $q \ll 1/d$, one can use the small momentum approximation, $U_p = U_0$, where this integral vanishes (see Appendix).

The remaining term in (\ref{vertex singular}), upon the regularization $U_p \to U_p e^{-p/\Lambda}$ is only logarithmic: $\int_0^\infty dp U_p =e^2 \ln (\Lambda d)$. The cut-off dependent part of the vertex correction is thus
\begin{equation}
\label{Kubo short range vertex}
(\Pi_{xx}^{(1)})_{V-AA1} = -\frac{N \omega^2 \delta v}{8 v_{0}^3 \sqrt{q^2v_{0}^2-\omega^2}}.
\end{equation}
The total cut-off dependent portion of the current-current correlation function is the sum of the self-energy (\ref{self-energy Kubo})  and the vertex correction (\ref{Kubo short range vertex}):
\begin{equation}
\label{Vertex Conductivity singular}
(\Pi_{xx}^{(1)})_{\Lambda}=  \frac{ N\omega^2 q^2  \delta v }{16 v_{0} (q^2v_{0}^2-\omega^2)^{3/2}}.
\end{equation}
This expression  matches exactly the $\Lambda$-dependent part (proportional to $\delta v$) of the polarization function correction, Eq.~(\ref{Polarization self-energy correction}), as both corrections satisfy the charge conservation condition (\ref{Ward identity1}).

We now further demonstrate that the cut-off independent first order corrections also separately satisfy the charge conservation condition.

Let us start with the second term in the parenthesis in Eq.~(\ref{Ver_cond3}) that gives rise to a finite integral that converges over small momenta $p, p' \sim q, \omega/v_0 \ll 1/d$. In this integral, it is sufficient to approximate,
 $U_{|{\bf p-p'}|} \approx U_0$. The ensuing momentum integrals are the same as in Eq.~(\ref{zero-order conductivity}):
\begin{align}
\label{Ver_cond4}
(\Pi_{xx}^{(1)})_{V-AA2} = - N\omega^4 U_0 \left(\sum_{\bf{p}} \frac{A_{\bf p}}{\Omega_{\bf p} (\omega^2-\Omega_{\bf p}^2)} \right)^2 \nonumber\\
=-\frac{N\omega^4 U_0}{(16v_0^2)^2 (\sqrt{q^2v_0^2-\omega^2})^2}.
\end{align}

The remaining two contributions into the vertex correction (\ref{Ver_cond2}) are likewise convergent. It is, therefore, possible to set in them $U_{|{\bf p}-{\bf p'}|}=U_0$. The contribution from the
term $-\omega^2 A_{\bf p}A_{{\bf p}'}$ in the numerator is
\begin{align}
\label{Ver_cond5}
(\Pi_{xx}^{(1)})_{V-AA3} = \frac{1}{2} N\omega^4 U_0 \left(\sum_{\bf{p}} \frac{A_{\bf p}}{\Omega_{\bf p} (\omega^2-\Omega_{\bf p}^2)} \right)^2 \nonumber\\
=\frac{N\omega^4 U_0}{2(16v_0^2)^2 (\sqrt{q^2v_0^2-\omega^2})^2}=-\frac{1}{2} (\Pi_{xx}^{(1)})_{V-AA2}.
\end{align}
Finally, the contribution from the $ B_{\bf p}B_{{\bf p}'}$ term is (see Appendix)
\begin{align}
\label{Ver_cond6}
(\Pi_{xx}^{(1)})_{V-BB} =- \frac{1}{2} N\omega^2 U_0  \left(\sum_{\bf{p}} \frac{B_{\bf p}}{\omega^2-\Omega_{\bf p}^2} \right)^2 \nonumber\\
							= -\frac{N\omega^2 U_0q^2}{2(16v_0)^2 (\sqrt{q^2v_0^2-\omega^2})^2} = \frac{q^2v_{0}^2}{2\omega^2} (\Pi_{xx}^{(1)})_{V-AA2}.
\end{align}
The three contributions (\ref{Ver_cond4})--(\ref{Ver_cond6}) add up to the cut-off independent portion of the current-current correlator,
\begin{equation}
\label{Corrections}
(\Pi_{xx}^{(1)})_{\text{non-}\Lambda} = -\frac{NU_{0} \omega^2(\omega^2+v_{0}^2q^2)}{2(16v_{0})^2v_{0}^2(\sqrt{q^2v_{0}^2-\omega^2})^2}.
\end{equation}
One can see that the cut-off independent part of the polarization function, Eq.~(\ref{Polarization vertex correction}), and the cut-off independent part of the current-current correlator (\ref{Corrections}) satisfy the charge conservation condition,
Eq. (\ref{Ward identity1}). Accordingly, we have demonstrated that the first order perturbation theory is consistent with the charge conservation.

Combining the self-energy and the vertex corrections to the longitudinal conductivity, we can write using identity (\ref{Kubo formula}), to the first order in the interaction:
\begin{equation}
\sigma^{(1)}_{xx}(\omega,{\bf q}) =  \frac{i e^2 N\omega q^2 v_{0}^2 \delta v }{16 v_{0} (q^2v_{0}^2-\omega^2)^{3/2}} - \frac{ie^2 NU_{0} \omega(\omega^2+v_{0}^2q^2)}{2(16v_{0})^2(q^2v_{0}^2-\omega^2)}.
\end{equation}
The first (cut-off dependent) term arises from large electron momenta $1/d <p < \Lambda$ and is described by Eq.~(\ref{Vertex Conductivity singular}). The second (cut-off independent) term comes  from small electron momenta $p \sim \omega/v_0,\,q$ and is described by Eq.~(\ref{Corrections}). The interaction correction is purely imaginary outside the electron-hole continuum, $qv_0>\omega$, but has a real part for $qv_0<\omega$ where generation of real electron-hole pairs is possible. In the short-range case, only the small momenta contribute to the interaction correction since $\delta v =0$.

In the limit of $q = 0$, the conductivity cannot distinguish between the screened Coulomb and the short-range interactions, and the interaction correction becomes,
\begin{equation}
\label{homog_cond}
\sigma^{(1)}_{xx}(\omega, 0) = \frac{ie^2 NU_{0} \omega}{2(16v_{0})^2}.
\end{equation}
Up to the overall sign, the correction (\ref{homog_cond}) coincides with the result of Ref.~\onlinecite{JVH}.

\section{The Bethe-Salpeter equation}
The polarization function of the non-interacting electrons, Eq.~(\ref{Non-interacting polarization}), displays the $1/\sqrt{qv_{0} -\omega}$ singularity near the boundary of the electron-hole continuum, $\omega=qv_{0}$. The degree of this singularity increases in the first-order interaction corrections. In the case of  screened Coulomb interaction, the strongest divergence appears in the self-energy correction, Eq.~(\ref{Polarization self-energy correction}), where it amounts to a simple shift of the boundary corresponding to the renormalization of the electron velocity, $v_0 \to v = v_0 +\delta v$. However, even after the main singularity is removed, the divergence $\sim 1/(qv_{0} -\omega)$ remains in the vertex correction,  Eq.~(\ref{Polarization vertex correction}). This indicates that for any small interaction $U_0$ (either screened Coulomb or short-range), the higher order interaction corrections become important in the proximity of the (renormalized) boundary of the electron-hole continuum, $\omega=qv$. Physically, this singular behavior is the result of interaction between an electron and a hole created upon absorption of the external field with frequency $\omega$, with the electron and the hole propagating almost parallel\cite{GFM} to the momentum ${\bf q}$ of the external field.

Summing up all the higher-order corrections is virtually an insurmountable task. We, therefore, focus on the subset of higher-order contributions with no intersection of the interaction lines (ladder diagrams) which are equivalent to the Bethe-Salpeter equation. Strictly speaking, the Bethe-Salpeter ladder, like other similar ``non-crossing'' approximations (such as the self-consistent Born approximation for graphene with disorder) is not well-controlled by any small parameter. Nonetheless, it represents an important starting point whose predictions could be tested by a comparison with numerical calculations.

In terms of the vertex function
$\hat \Gamma_{\omega, {\bf p},{\bf q}}$, illustrated in Fig.~2 in the ladder approximation, the polarization function is
 \begin{equation}
 \Pi(\omega,q) = -i \text{Tr} \sum_{{\bf p}}\int \frac{d\epsilon}{2\pi}
 			 \hat G_{\epsilon_+ \bf{p_+}}\hat \Gamma_{ \omega, {\bf p},{\bf q}} \hat G_{\epsilon_- {\bf p_-}}.
 \end{equation}
According to Fig. 2, the vertex function obeys the following integral equation:
 \begin{equation}
 \label{Gamma1}
 \hat \Gamma_{ \omega, {\bf p},{\bf q}} = 1 +i \sum_{{\bf k}}U_{{\bf p}-{\bf k}} \int \frac{d\epsilon}{2\pi}
 							 \hat G_{\epsilon_+ \bf{k_+}}\hat \Gamma_{ \omega, {\bf k},{\bf q}} \hat G_{\epsilon_- {\bf k_-}}.
 \end{equation}

 \subsection{Short-range interaction}

 Let us start with the case of the short-range interaction, where the interaction is strictly constant, $U_{{\bf p}-{\bf k}} =U_0$, and velocity renormalization is absent, $v =v_0$. In this case, dependence of the right-hand side of Eq.~(\ref{Gamma1}) on the momentum ${\bf p}$ disappears, and the vertex function depends only on the external $\omega$ and ${\bf q}$: $\hat \Gamma_{ \omega, {\bf p},{\bf q}} = \hat \Gamma_{ \omega,{\bf q}} $.
 From the isotropy of the system, we can conclude that the vertex function $\hat \Gamma_{ \omega,{\bf q}}$ depends on the direction of vector ${ \bf q}$ in the following way,
  \begin{figure}
 \includegraphics[width=6cm,height=6cm,keepaspectratio]{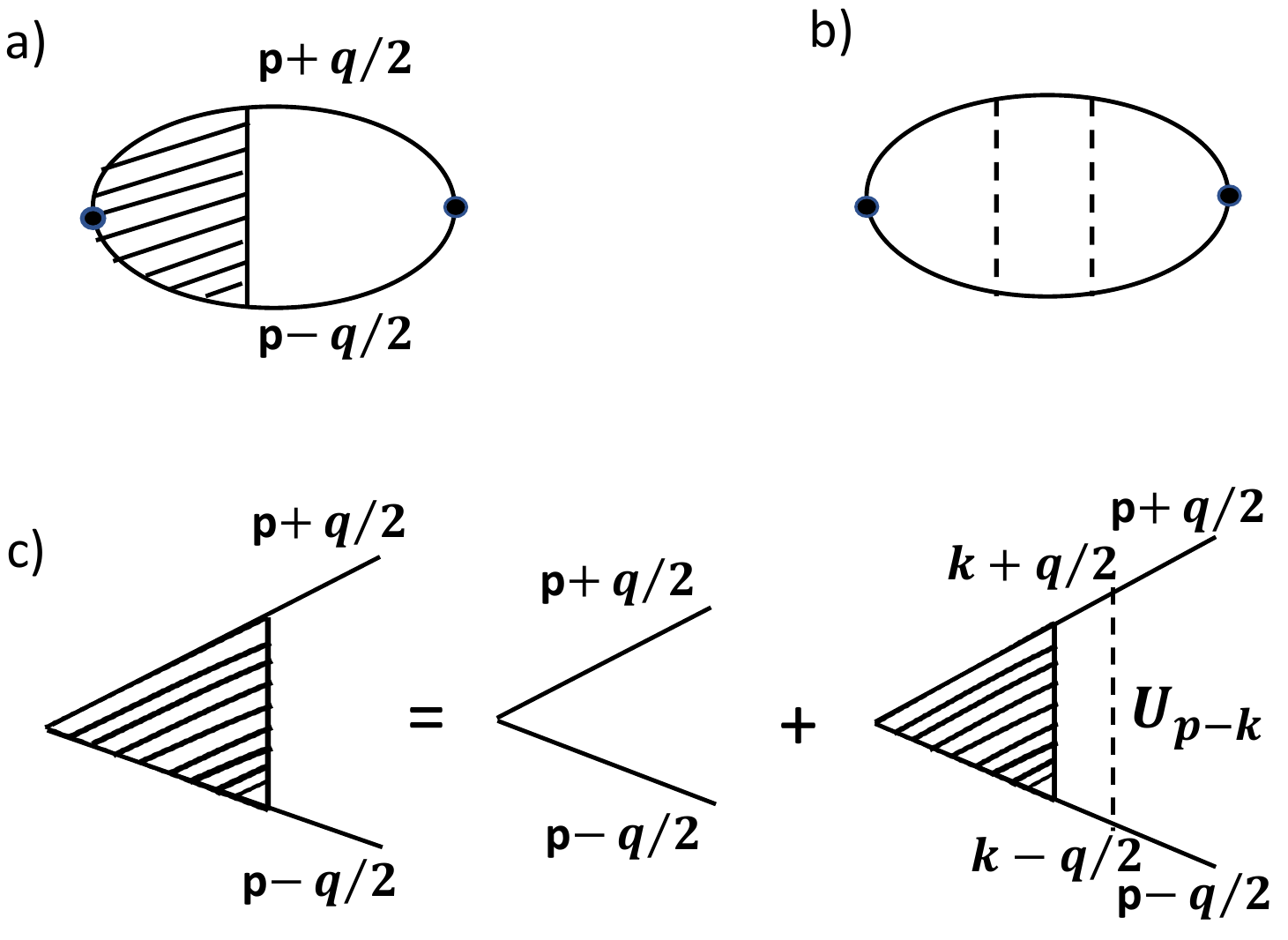}
 \caption{a) Total polarization function or conductivity corresponding to the vertices 1 or $ev_{0} \hat {\bf \sigma}$ respectively. b) Second order ladder diagram c) vertex summation to all orders in interaction.}
 \label{Fig. 2}
 \end{figure}
 \begin{equation}
 \label{Gamma2}
 \hat \Gamma_{ \omega,{\bf q}} = \Gamma_{0}(\omega,q) + \Gamma_{1}(\omega, q)\,{\hat \sigma}_{\bf q}.
 \end{equation}
Substitution of the anzats (\ref{Gamma2}) into Eq. (\ref{Gamma1}) leads to energy and momentum integrals. The first of those integrals is given by Eq.~(\ref{useful integral 1}):
 \begin{align}
\label{f}	
-i\sum_{\bf{k}} \int \frac{d\epsilon }{2\pi}\hat G_{\epsilon_+\, {\bf k}_+} \hat G_{\epsilon_- \,{\bf k}_-}&=f(\omega,q) +h (\omega,q){\hat \sigma}_{\bf q} \nonumber\\=& -\frac{q(qv+\omega \hat\sigma_{\bf q})}{32v \sqrt{q^2v^2-\omega^2}}.
\end{align}
 The second integral differs from the first by presence of the additional spin operator in the integrand:
\begin{align}
\label{g}																											
-i\sum_{\bf{k}}  \int \hat G_{\epsilon_+ \bf{k_+}} {\hat \sigma}_{\bf q} \hat G_{\epsilon_- {\bf k_-}}
										& =g(\omega,q) {\hat \sigma}_{\bf q}+ h(\omega,q),
\end{align}
 where the function $g(\omega,q)$ is substantially the current-current correlation function for the non-interacting electrons, see Eq.~(\ref{Kubo0}), $ \Pi_{xx}^{(0)}  (\omega, q) = 2Ng(\omega,q)$, and is given in Eq.~(\ref{zero-order conductivity}). Remember that the integral leading to the function $g(\omega,q)$ is formally divergent and should be regularized as described above (by performing the dimensional regularization or, equivalently, by subtracting the zero-frequency value $g(0,q)$). Note that this amounts to performing regularization in {\it every} instance where such diverging integral is encountered, which happens beginning from the terms of the second order in $U_0$. More specifically, to order $n$, the regularization must be carried out for $n-2$ internal rungs of the ladder, with the two outside rungs only resulting in the convergent expressions $f(\omega,q)$ and $h(\omega,q)$.

We now observe that the substitution of  Eq.~(\ref{Gamma2}) into Eq.~(\ref{Gamma1}) gives the matrix equation (arguments omitted),
 \begin{align}
\Gamma_{0} + {\hat \sigma}_{\bf q}\Gamma_{1}
 						= 1 - &U_0 (f +  h{\hat \sigma}_{\bf q}) \Gamma_{0}
 						 - U_0(g{\hat \sigma}_{\bf q} +h)\Gamma_{1},
 \end{align}
which amounts to two coupled scalar equations,
 \begin{eqnarray}
 \Gamma_{0}(1+U_0f) + \Gamma_{1}U_0 h &=& 1, \nonumber\\
 \Gamma_{0} U_0 h + \Gamma_{1} (1+U_0 g) &=& 0,
 \end{eqnarray}
 whose solutions are,
\begin{subequations}
 \begin{align}
 \Gamma_{0} =& \frac{1+U_0g}{(1+U_0f)(1+U_0g) -U_0^2 h^2} \\
  \Gamma_{1} =& -\frac{U_0h}{(1+U_0f)(1+U_0g) -U_0^2 h^2}.
 \end{align}
\end{subequations}
The functions encountered in these expressions satisfy the identity, $fg=h^2$, which follows from the actual expressions for these functions, determined, as explained above, by Eqs.~(\ref{useful integral 1}) and (\ref{Kubo0}).
  \begin{align}
 \label{Total polarization}
&\Pi(\omega,q)  = \frac{2N f}{1+U_0(f + g)} \nonumber\\
			&= -\frac{2Nq^2v ^2}{32v^2\sqrt{q^2v ^2 - \omega^2} - U_0(q^2v^2+\omega^2)}.
 \end{align}
In the absence of interactions, $U_{0} = 0$  we recover the non-interacting polarization function, Eq. (\ref{Non-interacting polarization}). The longitudinal conductivity now follows from Eq.~(\ref{Ward identity}). For example, the homogeneous longitudinal conductivity is
 \begin{equation}
 \label{Longitudinal_cond}
 \sigma_{\parallel}(\omega,0) = \frac{\sigma_0}{1- \displaystyle \frac{i U_0 \omega}{32v^2}}.
 \end{equation}
At $\omega \to 0$, the conductivity tends to its band value, $\sigma_0 =e^2N/16$. This can be anticipated from the fact that the interaction constant $U_0$ has the dimension of inverse mass, and thus one must use frequency to construct a (cut-off independent) dimensionless coupling strength, $ U_0 \omega /v^2$. Accordingly, in the limit of small frequencies, the interaction becomes negligible.

The polarization function (\ref{Total polarization}) has a pole at,
\begin{equation}
\label{resonance}
\omega^2 \approx q^2v^2\left[1-\left( \frac{U_0 q}{16v}\right)^2 \right].
\end{equation}
The pole signals that the response of the interacting electron system becomes resonant at frequencies determined by the last expression.

This expression holds provided that the interaction is sufficiently weak (so that the expression in the parenthesis is much smaller than one). In the limit of large interaction $U_0$, the zero-frequency response function (\ref{Total polarization})
predicts an instability at
\begin{equation}
q^* ={32 v}/{U_0}.
\end{equation}
The existence of such zero-frequency singularity would indicate that the ground state of the system is unstable with respect to formation of a charge-density wave.

\begin{figure}
 \includegraphics[width=8.5cm,height=7cm,keepaspectratio]{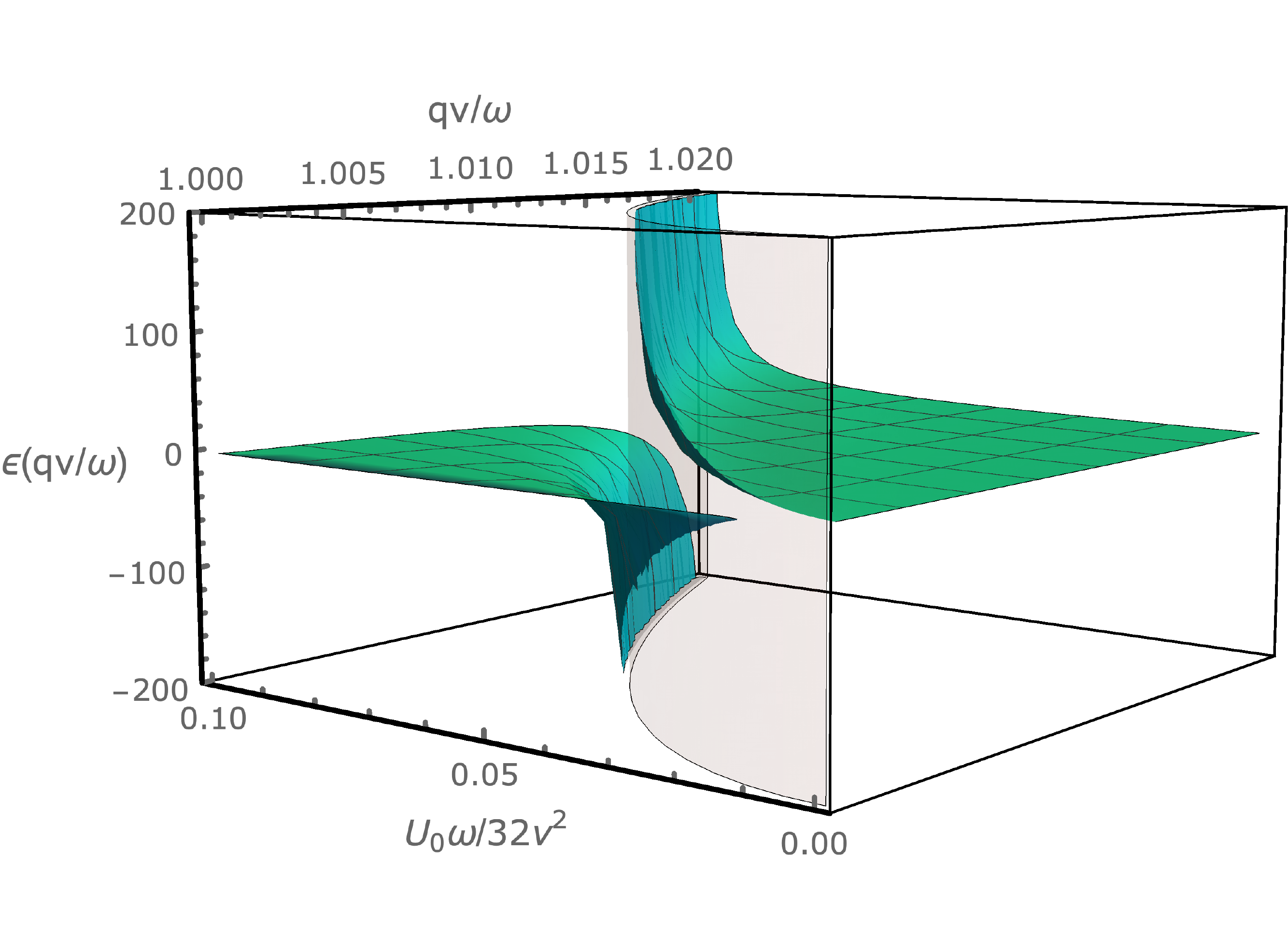}
 \caption{Plot of dielectric function, $\epsilon$ as a function of dimensionless parameters $qv/\omega$ and $U_{0}\omega/32v^2$.}
 \label{Fig. 3}
 \end{figure}

The polarization function $\epsilon(\omega,q) $ describes a static potential $V_{\rm tot}(\omega,q)$
occurring in the system in response to an external potential $V_{\rm ext}(\omega,q)$. This potential  $V_{\rm tot}$ acting in the system induces density variations $\rho (\omega,q)=\Pi (\omega,q) V_{\rm tot}$, which in turn are responsible for the induced part of the scalar potential, $V_{\rm tot}-V_{\rm ext} =U_0 \Pi (\omega,q) V_{\rm tot}$. This gives for the dielectric function, $\epsilon(\omega,q) = V_{\rm ext}/V_{\rm tot}=1-U_{0}\Pi(\omega,q)$. Fig.~\ref{Fig. 3} demonstrates  dependence of the dielectric function on $qv/\omega$ and dimensionless coupling strength $U_{0}\omega/32v^2$. The plot illustrates the existence of the resonance at the frequency determined  by Eq.~(\ref{resonance}).

\subsection{Screened Coulomb interaction}

Let us now turn to the case of the screened Coulomb interaction, where velocity renormalization $v_{0} \rightarrow v = v_{0} + \delta v $ is non-zero. As discussed above, $v$ is constant for low ($p \ll 1/d$) electron momenta,  $\delta v =e^2 \ln(\Lambda d)/4$, but
  has curvature for high momenta $p \gg 1/d$: $\delta v = e^2 \ln(\Lambda/p)/4$. Below we demonstrate that sufficiently close to the threshold $\omega =qv$, the contributions of the high electron momenta is small and the calculation of the polarization function is similar to that of the short-range interaction case.

  First, let us emphasize that at least for weak Coulomb interaction, the renormalization of the electron velocity can be taken to the first order. Indeed, as shown in Ref.~\onlinecite{Gonzales} for RPA diagrams and in Ref.~\onlinecite{EM} for crossing diagrams, to the second order in the (unscreened) Coulomb interaction, the velocity renormalization is $v=v_0+e^2 \ln(\Lambda/p)/4 +C (e^4/v_0)  \ln(\Lambda/p)$, with $C\sim 1$. In other words, no higher powers of the logarithms arise in the higher orders of the perturbation theory. Note that whereas for a suspended graphene the coupling constant $e^2/v_0$ is not small, weak couplings can be engineered by placing graphene on a substrate with a sufficiently high dielectric constant (which, effectively, reduces the value of the electric charge $e^2$).

To elucidate the role played by large momenta, consider the second-order ladder correction with two interaction lines.
\begin{align}
\label{second order vertex}
&\Pi^{(2)}_{V}(\omega,q) =  iN \text{Tr}\! \sum_{\bf{p},\bf{p'},\bf{p''}} U_{|{\bf p-p'}|}\int \frac{d\epsilon }{2\pi}\hat G_{\epsilon_+ \, {\bf p}_+} \hat G_{\epsilon_- \,{\bf p}_-} \nonumber\\
&\times U_{|{\bf p'-p''}|}\int \frac{d\epsilon' }{2\pi} \hat G_{\epsilon'_-\, {\bf p'}_- }\! \int \frac{d\epsilon'' }{2\pi} \hat G_{\epsilon''_-\, {\bf p''}_- } \hat G_{\epsilon''_+ \,{\bf p''}_+} \hat G_{\epsilon'_+ \,{\bf p'}_+}
\end{align}
The energy integrals over $\epsilon$ and $\epsilon''$, corresponding to the outside rungs, are the same as in Eq.~(\ref{useful integral}). It then follows (since they decrease sufficiently fast for large $p,\,p' \gg q$) that the remaining momentum integrals over ${\bf p}$ and ${\bf p}''$ converge. The integral over ${\bf p}'$, on the other hand, can extend to much larger momenta. The $1/q$-tail of the screened Coulomb interaction (in contrast to the short-range interaction) makes the ${\bf p}'$ integral convergent. Indeed, for large $p' \gg p,p''$ one can approximate, $U_{|{\bf p-p'}|}U_{|{\bf p'-p''}|} \approx U_{p'}^2$. The large momentum contribution is then encountered in the integral  that is similar to the integral in
Eq. (\ref{kubo zero order}) with the exception of the extra factor, $U_{p'}^2$:
\begin{align}
\label{screened calculation}
& \sum_{\bf p'} U_{p'}^2\frac{\Omega_{\bf p'} A_{\bf p'}}{\omega^2-\Omega_{\bf p'}^2} \nonumber\\
			&= \frac{-1}{4\pi v}\int\limits_0^\infty dp' \,U_{p'}^2 + \omega^2U_{0}^2\sum_{\bf p'} \frac{A_{\bf p'}}{\Omega_{\bf p'}(\omega^2-\Omega_{\bf p'}^2)},
\end{align}
where in the last line we separated the $\omega=0$ contribution. Using Eq.~(\ref{screened interaction}), we obtain that the first integral is a constant, $\int_0^\infty dp' \,U_{p'}^2 = 2U_{0}^2(\ln 2)/d$. The second integral amounts to the function $g(\omega,q)$. As a result, the expression in the right-hand side of (\ref{screened calculation}) becomes,
\begin{equation}
\frac{-U_{0}^2}{2\pi vd}\ln 2 -\frac{ \omega^2U_{0}^2}{16v^2\sqrt{q^2v^2-\omega^2}}.
\end{equation}
The first term is small compared with the second term (and hence, the screened Coulomb case maps on the short-range interaction) provided that $(\omega - qv_{0})/qv_{0} \ll (qd)^2$. As can be easily verified, the resonance (\ref{resonance}) falls within this interval if the interaction coupling constant is weak, $e^2/v \ll 1$.

\subsection{The transverse conductivity}
Turning now to the transverse conductivity $\sigma_\perp(\omega, q)$, we can write the pseudo-spin density correlation function,
 \begin{equation}
 \label{Transverse_cond}
 \Pi_{yy}(\omega,{\bf q}) = -i \text{Tr} \sum_{{\bf p}}\int \frac{d\epsilon}{2\pi}
 			 \hat G_{\epsilon_+ \bf{p_+}}\hat \Gamma_{\omega, {\bf p},{\bf q}} \hat G_{\epsilon_- {\bf p_-}} \hat \sigma_{y},
 \end{equation}
in terms of the vertex function $\hat \Gamma_{\omega, {\bf p},{\bf q}}$, shown in the ladder approximation in the same Fig.~2. The corresponding equation,
 \begin{equation}
 \label{Gamma1_cond}
 \hat \Gamma_{ \omega, {\bf p},{\bf q}} = \hat \sigma_{y} +i \sum_{{\bf k}}U_{{\bf p}-{\bf k}} \int \frac{d\epsilon}{2\pi}
 							 \hat G_{\epsilon_+ \bf{k_+}}\hat \Gamma_{ \omega, {\bf k},{\bf q}} \hat G_{\epsilon_- {\bf k_-}},
 \end{equation}
differs from Eq.~(\ref{Gamma1}) by the presence of the pseudospin operator $\sigma_y$.

As before, for the short-range interaction the right-hand side of Eq.~(\ref{Gamma1_cond}) does not depend on the momentum ${\bf p}$, and the vertex function depends only on the external $\omega$ and ${q}$: $\hat \Gamma_{ \omega, {\bf p},{q}} = \hat \Gamma_{ \omega,{q}} $.
As can be readily verified, the vertex function $\hat \Gamma_{ \omega,{q}}$  reduces to a single scalar function:
 \begin{equation}
 \label{Gamma2_cond}
 \hat \Gamma_{ \omega,{q}} = \Gamma_{2}(\omega, q)\,{\hat \sigma}_{y}.
 \end{equation}
Substitution of Eq.~(\ref{Gamma2_cond}) into Eq. (\ref{Gamma1_cond}) leads to the following energy and momentum integral, \begin{align}																										
-i\sum_{\bf{k}}  \int \hat G_{\epsilon_+ \bf{k_+}} {\hat \sigma}_{y} \hat G_{\epsilon_- {\bf k_-}}
										& =k(\omega,q) {\hat \sigma}_{y},
\end{align}
where the function $k(\omega,q)$ is substantially the transverse current-current correlation function for the non-interacting electrons, $ \Pi_{yy}^{(0)}  (\omega, q) = 2Nk(\omega,q)$, which is calculated  in the Appendix,
\begin{equation}
\label{function k}
k (\omega,{q}) = \frac{\sqrt{q^2v^2 - \omega^2}}{32v^2}.
\end{equation}
Like the integral in $g(\omega,q)$, the integral in Eq.~(\ref{Gamma2_cond}) is  formally divergent and should be regularized. The regularization can be
performed by subtracting the zero frequency and  momentum value, $k (\omega,{q}) \to k (\omega,{q})-k(0,0)$, which results in the value (\ref{function k}).

Interestingly, unlike the longitudinal conductivity, see function $g(\omega,q)$ for which the same result is obtained regardless of whether the value $g(0,q)$ or $g(0,0)$ is subtracted (as the difference between these two values is zero),
for the transverse conductivity $k(0,q) \ne k(0,0)$. Accordingly, the regularization procedure $k (\omega,{ q}) \to k (\omega,{q})-k(0,q)$ leads to a result that is different\cite{regularization} from Eq.~(\ref{function k}).
For the transverse conductivity, there is no charge conservation (as the induced charge is zero) to guide one in selecting the right regularization procedure.

We notice, however, that the latter regularization leads to the transverse conductivity having a negative imaginary part for $qv>\omega$, which would indicate an {\it inductive} impedance of the system -- a situation that appears rather unphysical in the absence of induced magnetic fields.

Using Eq.~(\ref{function k}), we obtain from Eq.~(\ref{Gamma1_cond}) that
\begin{equation}
\Gamma_{2}(\omega,q) = \frac{1}{1+U_{0} k(\omega,q) }
\end{equation}
This gives the transverse conductivity in the Bethe-Salpeter approximation,
 \begin{align}
 \label{transverse result}
\sigma_\perp(\omega,{q}) &= \frac{2ie^2 v^2 N}{\omega} \frac{k(\omega,q)}{1+U_{0} k(\omega,q)} \nonumber\\
 					&= \frac{2ie^2 v^2 N \sqrt{q^2v ^2 - \omega^2}}{\omega (32v^2 + U_{0}\sqrt{q^2v ^2 - \omega^2})}.
 \end{align}

In the limit of $q=0$, the transverse conductivity (\ref{transverse result}) coincides with the longitudinal conductivity, Eq. (\ref{Longitudinal_cond}), as required by the isotropy of the system in the homogeneous limit.

At large momenta $qv\gg \omega$, the transverse conductivity (\ref{transverse result}) tends to a constant $\sigma_\perp \to 2ie^2 v^2 N/ (\omega U_0)$ that depends on the interaction $U_0$. (The longitudinal conductivity vanishes in the same limit.)

\section{Summary and Conclusions}
We have calculated the first order corrections to the polarization function $\Pi(\omega,q)$ and the longitudinal conductivity $\sigma(\omega,q)$ for a screened Coulomb interactions and compared them to the corresponding quantities in the short-range interaction model. We have verified that divergent integrals are regularized (using dimensional regularization or by subtracting the $\omega=0$ contributions from divergent integrals) in a way that ensures charge conservation (the Ward identity). Based on the  understanding of the first order perturbation theory, we have solved the Bethe-Salpeter equation for the polarization operator to collect the increasingly singular (near $\omega = qv$) perturbative corrections. Sufficiently close to the boundary of the electron-hole continuum, the screened Coulomb interaction leads to the same results as the short-range interaction model, with one exception that in the former case the boundary is shifted as a result of the interaction-induced renormalization of the electron velocity.

These findings predict a resonant response of interacting electron-hole pairs below the threshold $qv=\omega$, at the frequency (\ref{resonance}),  and further predict an instability (revealed in the zero frequency response) for sufficiently strong interactions. It should be emphasized that the Bethe-Salpeter approach is somewhat uncontrolled, as it neglects crossing diagrams (diagrams that describe virtual processes having multiple electron-hole pairs in the intermediate states). Although a natural starting point, this approach remains to be tested further (e.g., via numerical simulations).

\section{Acknowledgements}
Numerous valuable discussions with O.A. Starykh are gratefully acknowledged.
The work is supported by DOE, Office of Basic Energy Sciences, Grant No.~DE-FG02-06ER46313. %OS is supported by NSF DMR grant No.~1928919.

\appendix
\section{Calculation of integrals}
% \section{Logarithmic integral in the self-energy}
% The velocity renormalization (\ref{velocity renormalization 3}) involves the following integral that, upon adding and subtracting the integral $\int\limits_1^{\Lambda d} dx/x$, is transformed as follows ($x=kd$),
% \begin{align}
%\label{velocity renormalization 4}
%I_1 =&-1 + \int\limits_0^{\Lambda d} \frac{dx}{x} \left( 1- e^{-x} \right) \nonumber\\
%			   & = \ln{(\Lambda d)}-1 +\int\limits_0^{1} \frac{dx}{x} - \int\limits_0^{\Lambda d} \frac{dx}{x}e^{-x}.
%\end{align}
%In the last integral it is now possible to take the upper limit to infinity ($\Lambda d \to \infty$) to obtain,
 %\begin{align}
%\label{velocity renormalization 5}
%I_1 =& \ln{(\Lambda d)}-1 +\int\limits_0^{1} \frac{dx}{x} - \int\limits_0^{\infty} \frac{dx}{x}e^{-x} \nonumber\\
%			   & = \ln{(\Lambda d)}-1+\gamma = \ln{(\Lambda d \,c_1)},
%\end{align}
%where $\gamma =0.577$ is the Euler-Macheroni constant.

% \subsection{APPENDIX A: CALCULATION OF INTEGRALS}
(i) To carry out the momentum integral (see  Eq.~({\ref{useful integral 1}}) where $v_0$ is replaced with $v$),
\begin{align}
\label{First integral1}
I_{1} =& \sum_{\bf{p}} \int \frac{d\epsilon }{2\pi}\hat G_{\epsilon_+\, {\bf p}_+} \hat G_{\epsilon_- \,{\bf p}_-} \nonumber \\ &= i\sum_{\bf{p}}\frac{ v (p_++p_-) (1-\hat\sigma_{{\bf p}_+} \hat\sigma_{{\bf p}_-}) +\omega(\hat\sigma_{{\bf p}_+} - \hat\sigma_{{\bf p}_-}) }{2[\omega^2-v^2(p_++p_-)^2]},
\end{align}
it is convenient to shift momentum, ${\bf p} \rightarrow {\bf p+q/2}$, so that $1-\hat\sigma_{{\bf p}_+} \hat\sigma_{{\bf p}_-} \rightarrow 1-\hat\sigma_{{\bf p+q}} \hat\sigma_{{\bf p}}$ and $\hat\sigma_{{\bf p}_+} - \hat\sigma_{{\bf p}_-} \rightarrow \hat\sigma_{{\bf p+q}} - \hat\sigma_{{\bf p}}$. Using  $\hat\sigma_{\bf p} = \hat\sigma_{\bf q } \cos\theta_{\bf p} + \sigma_{{\bf z} \times {\bf q}}\sin \theta_{\bf p} $ (where ${\bf z}$ is the direction perpendicular to the plane of graphene and $\theta_{\bf p}$, as before, is the angle that the direction of ${\bf p}$ makes with the vector ${\bf q}$) and noticing that the integrals of the sine terms vanish by symmetry, we can write,
\begin{align}
\label{First integral2}
I_{1} = i\sum_{{\bf p}} &\frac{1}{2[\omega^2 - v^2 (p+|{\bf p+q}|)^2]}\bigg(v(p+|{\bf p+q}|) \nonumber\\
	&\times (1-\cos{\theta_{{\bf p,p+q}}})+ \omega \sigma_{\bf q} (\cos{\theta_{{\bf p+q}}} - \cos{\theta_{{\bf p}}}) \bigg),
\end{align}
where $\theta_{{\bf p,p+q}}$ is the angle between vectors ${\bf p}$ and ${\bf p}+{\bf q}$.
To calculate the integral, we  use the absolute values $p$ and $ k =|{\bf p}+{\bf q}|$ as new variables, to replace the angle variable $\theta_{\bf p}$ in the integral over $d^2p =p \,dp \,d\theta_{\bf p}$ with  $k$. To make use of the new variables, we first notice the identity,
\begin{equation}
\label{Identity}
\int\limits_0^\infty dk\frac{k}{pq} \delta\Big(\frac{k^2 - p^2 - q^2}{2pq} -\cos \theta_{\bf p} \Big) = 1,
\end{equation}
and express cosines of the encountered angles via $p$ and $k$:
\begin{eqnarray}
\label{Identities}
1 - \cos{\theta_{{\bf p,p+q}}} &=& \frac{p+q \cos\theta_{\bf p}}{k} = \frac{q^2 - (k-p)^2}{2kp}, \nonumber\\
\cos{\theta_{{\bf p+q}}} - \cos{\theta_{{\bf p}}} &=& \frac{p\cos{\theta_{\bf p}} + q}{k}  - \cos{\theta_{\bf p}} \nonumber\\
				=  &&\frac{(k+p)(q^2-(k-p)^2)}{2pkq}.
\end{eqnarray}
The  integral over $d \theta_{{\bf p}}$  now leads to the following expression,
\begin{align}
\int\limits_{0}^{2 \pi}  d\theta_{\bf p} & \delta\Big(\frac{k^2 - p^2 - q^2}{2pq} -\cos \theta_{\bf p} \Big) \nonumber\\
			&= \frac{4pq}{\sqrt{((k+p)^2 - q^2)(q^2-(k-p)^2)}}.
\end{align}
%\begin{align}
%I_{1} = & \sum_{{\bf p}} \bigg( \frac{q^2 - (k-p)^2}{2pk}\bigg)\bigg(\frac{v_{0}(p+k)+ \omega\sigma_{x}(k+p)/q}{\omega - \beta v_{0} (p+|{\bf p+q}|)}\bigg) \nonumber\\
%	 = &\frac{(vq + \sigma_{x}\omega)}{2\pi^2q}\int\frac{dp dk}{\sqrt{((k+p)^2 - q^2)(q^2-(k-p)^2)}} \nonumber\\
%		& ~~~~~~~~~~~\times \frac{(q^2 - (k-p)^2)(k+p)}{(\omega^2-v_{0}^2(p+k)^2)}
%\end{align}
Because in the remaining integrals over $dp\,dk$ in Eq. (\ref{First integral2}) the integrand depends only on the sum $k+p$ and the difference $k-p$, this suggests rotating the
 integration variables, $k+p=qx $, and $ k-p=qy$, to factorize the integrals as follows,
\begin{align}
\label{int}
I_{1} = &\frac{i q^2 (vq+\omega \sigma_{\bf q})}{2\pi^2}\int_{1}^{\infty} \frac{dx~x}{\sqrt{x^2-1}\,(\omega^2- q^2v^2x^2)} \nonumber\\
	&\times \int_{-1}^{1}dy\sqrt{1-y^2} = -\frac{iq(qv+\omega \hat\sigma_{\bf q})}{32v\sqrt{q^2v^2-\omega^2}}.
\end{align}

(ii) To calculate the momentum integral in Eq. (\ref{vertex singular}) ($v_{0}$ is replaced with $v$),
\begin{equation}
I_{2} = \sum_{\bf{p}} \left( \frac{A_{\bf p} }{\Omega_{\bf p}} -\frac{1}{2vp}\right)U_p,
\end{equation}
where $A_{\bf p}= 1-\cos(\theta_{{\bf p}_+}+\theta_{{\bf p}_-})$ and $\Omega_{{\bf p}} = v(p_++p_-)$, we shift momentum, ${\bf p} \rightarrow {\bf p+q/2}$, and rewrite the above integral as,
\begin{equation}
I_{2} =  \sum_{\bf p} \left(\frac{1-\cos(\theta_{{\bf p}}+\theta_{{\bf p+q}})}{v(p+|{\bf p+q}|)} -\frac{1}{2vp}\right)U_p.
\end{equation}
As above, we use the absolute values $p$ and $ k =|{\bf p}+{\bf q}|$ as new variables, to replace the angle variable $\theta_{\bf p}$ in the integral over $d^2p =p \,dp \,d\theta_{\bf p}$ with  $k$. The cosine in the integrand is simplified as, $1-\cos(\theta_{{\bf p}}+\theta_{{\bf p+q}}) = (k+p)^2(q^2-(k-p)^2)/2pkq^2$ and using the identity in Eq. (\ref{Identity}) gives,
\begin{align}
I_{2} =  \frac{U_{0}}{2\pi^2q^2v}\int dp~dk &\frac{(k+p)\sqrt{q^2-(k-p)^2}}{\sqrt{(k+p)^2-q^2}} \nonumber\\
		& -\int dp \frac{U_{0}}{4\pi v}.
\end{align}
In the last line we noticed that only small values of momentum ${\bf p}$ are important to the integral where $U(p) \approx U_{0}$. \footnote{The corrections coming from the momentum dependence of interaction will be proportional to $qd \ln(1/(qd))$, and is thus small for $qd \ll 1$.}
The first integral is simplified by rotating the integration variables, $k+p = qx$ and $k-p = qy$,
\begin{equation}
I_{2} =  \frac{U_{0}}{4\pi v}\left (\frac{q}{2}\int_{1}^{2\Lambda/q} dx\frac{x}{\sqrt{x^2-1}}  -\int_0^{\Lambda} dp\right) = 0.
\end{equation}
The integrals in the parenthesis exactly cancel each other and gives zero.

(iii) To calculate the integral in Eq. (\ref{zero-order conductivity}) and Eq. (\ref{Ver_cond4})($v_{0}$ is replaced with $v$),
\begin{equation}
I_{3} =  \sum_{\bf p} \frac{ A_{\bf p}}{\Omega_{\bf p}(\omega^2-\Omega_{\bf p}^2)},
\end{equation}
where $A_{\bf p}= 1-\cos(\theta_{{\bf p}_+}+\theta_{{\bf p}_-})$, and $\Omega_{{\bf p}} = v(p_++p_-)$, we shift momentum, ${\bf p} \rightarrow {\bf p+q/2}$, and rewrite the above integral as,
\begin{equation}
I_{3} =  \sum_{\bf p} \frac{1-\cos(\theta_{{\bf p}}+\theta_{{\bf p+q}})}{v(p+|{\bf p+q}|)(\omega^2-v^2(p+|{\bf p+q}|)^2))}.
\end{equation}
As before, the encountered cosines in the integrand are expressed via $p$ and $k = |{\bf p+q}|$, and using the identity in Eq. (\ref{Identity}), $1-\cos(\theta_{{\bf p}}+\theta_{{\bf p+q}}) = (k+p)^2(q^2-(k-p)^2)/2pkq^2$,
\begin{equation}
I_{3} =  \frac{1}{2\pi^2q^2v}\int dp~dk \frac{(k+p)\sqrt{q^2-(k-p)^2}}{\sqrt{(k+p)^2-q^2}(\omega^2-v^2(p+k)^2))}.
\end{equation}
Rotating the integration variables, $k+p = qx$ and $k-p = qy$ (see Eq. (\ref{int})),
\begin{equation}
I_{3} = -\frac{1}{16v^2\sqrt{q^2v^2-\omega^2}}.
\end{equation}

(iv) To calculate the integral in Eq. (\ref{Ver_cond6}) ($v_{0}$ is replaced with $v$),
\begin{align}
I_{4} = \sum_{\bf{p}} \frac{B_{\bf p}}{\omega^2-\Omega_{\bf p}^2},
\end{align}
where $B_{\bf p}= \cos{\theta_{{\bf p_+}}} - \cos{\theta_{{\bf p_-}}}$ and $\Omega_{{\bf p}} = v(p_++p_-)$ we shift momentum, ${\bf p} \rightarrow {\bf p+q/2}$, and rewrite the above integral as,
\begin{equation}
I_{4} = \sum_{\bf{p}} \frac{\cos{\theta_{{\bf p+q}}} - \cos{\theta_{{\bf p}}}}{\omega^2-v^2(p+|{\bf p+q}|)^2}.
\end{equation}
From Eq. (\ref{Identities}), we see that the integral $I_4$ is related to $I_3$ so that $I_4 = qvI_3$,
\begin{align}
I_{4} = -\frac{q}{16v\sqrt{q^2v^2-\omega^2}}.
\end{align}

(v) To the zeroth order in the interaction, the transverse pseudo-spin density correlator is,
\begin{align}
k (\omega,{q}) = -\frac{i}{2} \text{Tr} \sum_{\bf p}\int \frac{d\epsilon }{2\pi}\hat G_{\epsilon_+\, {\bf p}_+} \hat\sigma_{y} \hat G_{\epsilon_- \,{\bf p}_-} \hat \sigma_y.
\end{align}
Taking the energy integral,
\begin{align}
&\int \frac{d\epsilon }{2\pi}\hat G_{\epsilon_+\, {\bf p}_+} \hat\sigma_{y} \hat G_{\epsilon_- \,{\bf p}_-}\nonumber \\
&= i\frac{\omega  (\hat\sigma_{{\bf p}_+}\hat\sigma_{y} - \hat\sigma_{y}\hat\sigma_{{\bf p}_-}) +\Omega_{\bf p}(\hat\sigma_{y}-\hat\sigma_{{\bf p}_+}\hat\sigma_{y} \hat\sigma_{{\bf p}_-})}{2(\omega^2-\Omega_{\bf p}^2)}.
\end{align}
Multiplying this expression by $\hat \sigma_y$ and taking the trace, we obtain,
\begin{equation}
k (\omega,{q}) = \frac{1}{2} \sum_{\bf p} \frac{\Omega_{\bf p} C_{\bf p}}{\omega^2-\Omega_{\bf p}^2},
\end{equation}
where $C_{\bf p}= 1+\cos(\theta_{{\bf p}_+}+\theta_{{\bf p}_-})$  and $\Omega_{{\bf p}} = v(p_++p_-)$. $\theta_{{\bf p}_\pm}$ denote the angles that the momenta ${\bf p}_\pm$ make with the direction of ${\bf q}$.
Using the same method as in the rest of this Appendix, namely, shifting the momentum, ${\bf p} \rightarrow {\bf p+q/2}$,  making use of the variables $p$ and $k = |{\bf p+q}|$, and then rotating them by $k+p = x$ and $k-p = y$,
we arrive at the following integral (note that $C_{\bf p} = (p-k)^2 [(p+k)^2 -q^2]/2pkq^2$),
\begin{equation}
\label{kubo zero order transverse}
k (\omega,{q}) = \frac{v}{16\pi}\int\limits_{q}^{\infty}\frac{dx~x\sqrt{x^2-q^2}}{(\omega^2-v^2x^2)}.
\end{equation}
The obtained integral is power-law divergent. To regularize this divergence we subtract from the integrand in Eq.~(\ref{kubo zero order transverse}) its zero-frequency and momentum value.
This is done in two steps, $k(\omega,q) \rightarrow k(\omega,q) - k(0,0)$ as follows: $(k(\omega,q) - k(0,q)) + (k(0,q) - k(0,0))$,
\begin{align}
k (\omega,{q}) \to &\frac{\omega^2}{16\pi v}\int\limits_{q}^{\infty}\frac{dx\sqrt{x^2-q^2}}{x(\omega^2-v^2x^2)} \nonumber\\
				&+\frac{1}{16\pi v}\bigg(\int\limits_0^\infty ~dx-\int\limits_q^\infty~dx\frac{\sqrt{x^2-q^2}}{x}\bigg).
\end{align}
The integral in the parenthesis gives $\pi/2$ so that the above expression reproduces Eq.~(\ref{function k}).


\begin{references}
\bibitem{PN} D. Pines and P. Nozieres, {\it The Theory of Quantum Liquids} (Benjamin, New York, 1966).

\bibitem{CN} A.~H.~Castro Neto,
F.~Guinea, N.~M.~R.~Peres, K.~S.~Novoselov, and A.~K.~Geim, Rev.
Mod. Phys. 81, 109 (2009).

\bibitem{Shung}K. W. Shung, Phys. Rev. B {\bf 34}, 1264 (1986).

\bibitem{Gonzales} J. Gonz\'alez, F. Guinea, and M.A.H. Vozmediano,
Nucl. Phys. B {\bf 73}, 125411 (1994); % J. Gonz\'alez, F. Guinea, and M.A.H. Vozmediano,
Phys. Rev. B {\bf 59}, R2474 (1999).

\bibitem{LFS} A. W. W. Ludwig, M. P. A. Fisher, R. Shankar, and
G. Grinstein, Phys. Rev. B {\bf 50}, 7526 (1994).

\bibitem{NBG} R. R. Nair, P. Blake, A. N. Grigorenko, K. S. Novoselov,
T. J. Booth, T. Stauber, N. M. R. Peres, and A. K.
Geim, Science {\bf 320}, 1308 (2008).

\bibitem{SS1} D. E. Sheehy and J. Schmalian, Phys. Rev. Lett. {\bf 99},
226803 (2007).

\bibitem{HJV} I. F. Herbut, V. Juricic, and O. Vafek, Phys. Rev. Lett.
{\bf 100}, 046403 (2008).

\bibitem{M} E. G. Mishchenko, Europhys. Lett. {\bf 83}, 17005 (2008).

\bibitem{SS2} D. E. Sheehy and J. Schmalian, Phys. Rev. B {\bf 80}, 193411
(2009).

\bibitem{AVP} S. H. Abedinpour, G. Vignale, A. Principi, M. Polini,
W.-K. Tse, and A. H. MacDonald, Phys. Rev. B {\bf 84},
045429 (2011).

\bibitem{SF} I. Sodemann and M. M. Fogler, Phys. Rev. B {\bf 86}, 115408
(2012).

\bibitem{TK} S. Teber and A. V. Kotikov, Europhys. Lett. {\bf 107}, 57001 (2014).

\bibitem{LOS} J. M. Link, P. P. Orth, D. E. Sheehy, J. Schmalian, Phys. Rev. B {\bf 93}, 235447 (2016).

\bibitem{JVH} V. Juricic, O. Vafek, and I. F. Herbut, Phys. Rev. B {\bf 82},
235402 (2010).

\bibitem{GMP} A. Giuliani, V. Mastropietro, and M. Porta, Phys. Rev. B {\bf 83}, 195401 (2011).

\bibitem{footnote1} For $qv_{0} < \omega$, the branch of the square root function should be selected (for positive $\omega$) as follows: $\sqrt{q^2v_{0}^2-\omega^2} \rightarrow -i \sqrt{\omega^2-q^2v_{0}^2}$.

\bibitem{TLR} Ho-Kin Tang, J. N. Leaw, J. N. B. Rodrigues, I. F. Herbut, P. Sengupta, F. F. Assaad, and  S. Adam, Science {\bf 361}, 570 (2018).

\bibitem{EM} E.G. Mishchenko, Phys. Rev. Lett. {\bf 98}, 216801 (2007).

\bibitem{GFM} S. Gangadharaiah, A. M. Farid, and E.G. Mishchenko, Phys. Rev. Lett. {\bf 100}, 166802, 2008.

\bibitem{regularization} The corresponding expression differs from Eq.~(\ref{function k}) by the replacement $\sqrt{q^2 v^2 - \omega^2} \to \sqrt{q^2 v^2 - \omega^2}-qv$.

\end{references}
\end{document}